\documentclass[10pt]{article}

\usepackage[margin=0.8in]{geometry}
\usepackage[T1]{fontenc}
\usepackage[utf8]{inputenc}
\usepackage{graphicx}
\usepackage{booktabs}
\usepackage{array}
\usepackage{tabularx}
\usepackage{enumitem}
\usepackage{xurl}
\usepackage[hidelinks]{hyperref}
\usepackage{listings}
\usepackage{caption}

\graphicspath{{./}}
\setlist[itemize]{leftmargin=*, itemsep=2pt, topsep=3pt}
\setlist[enumerate]{leftmargin=*, itemsep=2pt, topsep=3pt}
\title{PUDA: An AI-Native Hardware Harness for Self-Driving Laboratories}
\author{
Zekun Ren\textsuperscript{1,*},
Hongzhao Tan\textsuperscript{1,*},
JiaEn Yee\textsuperscript{1,2},
Kedar Hippalgaonkar\textsuperscript{1,2,*}\\[3pt]
\small \textsuperscript{1}Berkeley Education Alliance for Research in Singapore (BEARS)\\
\small \textsuperscript{2}School of Materials Science and Engineering, Nanyang Technological University (NTU)\\
\small \textsuperscript{*}Corresponding authors: Zekun Ren, Hongzhao Tan, and Kedar Hippalgaonkar\\
\small Correspondence: \href{mailto:danny.ren@bears-berkeley.sg}{danny.ren@bears-berkeley.sg}; \href{mailto:hongzhao.tan@bears-berkeley.sg}{hongzhao.tan@bears-berkeley.sg}; \href{mailto:kedar@ntu.edu.sg}{kedar@ntu.edu.sg}
}
\date{}

\begin{document}
\maketitle

\begin{abstract}
Physical Unified Device Architecture (PUDA) is an AI-native hardware harness for self-driving laboratories (SDLs). Rather than building a human-centered graphical user interface (GUI) orchestration layer, PUDA creates a command-line runtime environment that lets agents observe, orient, decide, and act over experiments while hardware execution remains deterministic, atomic, and auditable. Headless by design, devices appear through discoverable command-line interfaces, JSON protocols are routed through a distributed messaging system, and command responses, data products, and reports are preserved as structured records. PUDA organizes protocols, runs, samples, measurements, and command logs into an AI-native data structure linked by run identifiers and timestamps, preserving provenance from submitted protocol through hardware response to resulting data products. PUDA separates scientific orchestration from physical operation and data telemetry: agents choose experiments, while PUDA executes validated commands and captures provenance-linked state, responses, and data. The contribution is not another optimizer, orchestrator, or recipe language. It is a practical execution and data environment for agentic SDLs; the broader physical AI implication is that PUDA provides an AI-native hardware harness for AI systems to interact with physical tools.
\end{abstract}


\section{Introduction}
Self-driving laboratories (SDLs) combine automated instruments, experimental design algorithms, and machine-learning feedback loops to accelerate scientific discovery \cite{tom2024,abolhasani2023,hase2019}. In an SDL, software selects an experiment, hardware executes it, measurements are analyzed, and the result informs the next experiment. This closed-loop structure has enabled increasingly complex campaigns in chemistry, materials science, and biology.

However, many SDL systems still expose the laboratory through human-first workflow abstractions. A scientist or developer often designs a fixed workflow graph, writes bespoke scripts around a particular instrument stack, or encodes orchestration logic inside a GUI-driven platform. These approaches are useful for repeatable automation, but they become rigid when an AI agent must recover from errors, inspect intermediate state, revise the next protocol, or coordinate multiple devices during a live campaign. As agents become better at planning and tool use, the bottleneck shifts from ``can the workflow be automated?'' to ``can the laboratory expose a stable, inspectable runtime environment that an agent can operate?''

The difficulty is also a data-infrastructure problem. For decades, scientific communities have developed domain ontologies, common schemas, metadata standards, and repositories to make heterogeneous experimental records interpretable and reusable. The FAIR principles consolidate this agenda around data that are Findable, Accessible, Interoperable, and Reusable, with explicit emphasis on machine actionability and provenance \cite{wilkinson2016fair}. Yet in automated laboratories, protocols, device commands, machine state, measurements, and reports are often produced by separate systems without stable identifiers or a shared execution context, limiting both human reuse and autonomous-agent reasoning.

Physical Unified Device Architecture (PUDA) addresses these coupled interface and data-infrastructure problems. PUDA is an AI-native hardware harness for self-driving laboratories: a headless, command-line runtime environment that makes the laboratory accessible to AI agents. Instead of building another human-centered orchestration GUI, PUDA exposes the lab as an AI-first command environment. Agents interact with discoverable CLI commands, structured JSON protocols, machine state, logs, reports, and data products. PUDA also organizes protocols, runs, samples, measurements, and command logs as machine-readable records linked by stable identifiers and timestamps, preserving provenance from submitted protocol through hardware response to resulting data products. This AI-native data structure supports FAIR objectives and provides a common execution context in which domain-specific data and ontologies can be used; it does not replace those ontologies. Hardware execution remains deterministic, atomic, and auditable because physical actions are dispatched through engineer-authored driver methods rather than generated ad hoc during an experiment \cite{leong2022framework}.

This design lets an AI agent complete an Observe, Orient, Decide, and Act (OODA) loop over a physical laboratory. The agent observes command outputs, machine state, logs, videos, and reports; orients by interpreting those records against the experiment objective; decides the next protocol or recovery action; and acts by issuing validated PUDA commands. PUDA does not prescribe the agent's reasoning strategy. It supplies the stable command, transport, and record layer that makes agent-led reasoning executable in the physical world.

At the transport layer, PUDA uses a general-purpose distributed messaging system rather than a lab-specific point-to-point integration layer. PUDA selects NATS.io as a lightweight, high-performance publish/subscribe messaging system for hardware services, command dispatch, telemetry, and state updates \cite{natsdocs}. JetStream extends this messaging layer with durable streams and queues, allowing commands, responses, and experiment events to be retained and replayed. This gives PUDA a shared communication backbone for both real-time operation and post-run auditability.

PUDA uses this layered architecture to make agentic hardware operation safer and recoverable. Edge services expose only reviewed driver methods, so an agent can request an action but cannot rewrite device-control code during a run. Before any command reaches hardware, PUDA checks that the requested operation uses a known command name, required inputs, valid parameter types, and allowed value ranges. When something fails, progressive discovery, CLI error handling, and command navigation give the agent a clear next step instead of a silent crash. Because PUDA routes commands and telemetry through NATS/JetStream, many devices and agents can operate at the same time: each device listens on its own command channel, agents can monitor state or analyze outputs in parallel, and durable queues keep long-running tasks coordinated without forcing the whole campaign into one fixed workflow graph. This design extends naturally to distributed campaigns across multiple tools, laboratories, and agent roles, where execution, monitoring, analysis, and planning can proceed concurrently while sharing the same auditable record.

\begin{figure}[t]
\centering
\includegraphics[width=0.92\linewidth]{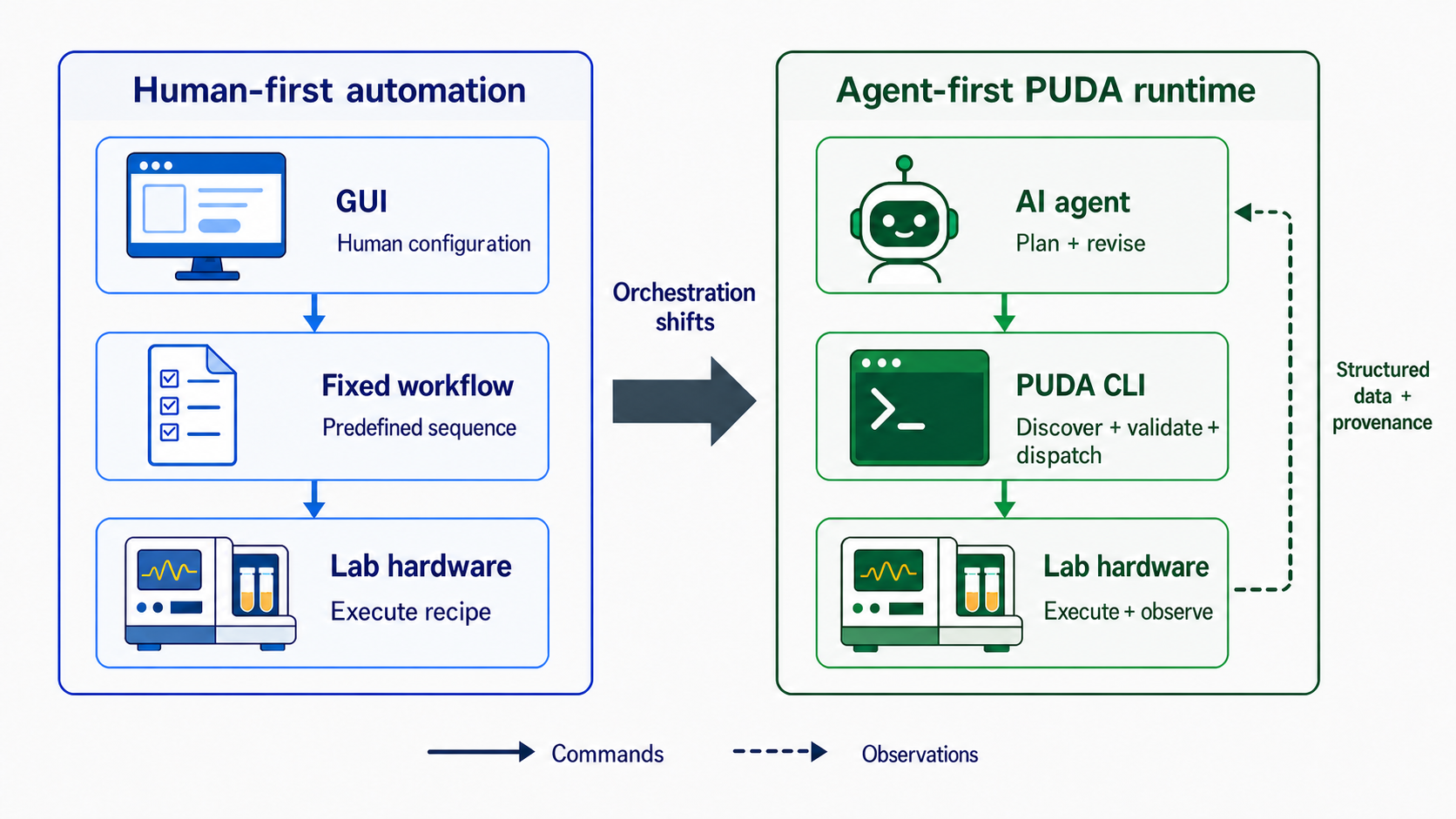}
\caption{PUDA changes where orchestration lives. Human-first automation fixes workflows before a run; PUDA keeps the lab interface stable while an agent revises protocols between runs.}
\label{fig:paradigm}
\end{figure}

The contribution of PUDA is therefore not another optimizer, orchestrator, or recipe language. PUDA standardizes the interface where AI agents interact with the laboratory. It provides a practical execution and data environment in which agents can operate SDLs while hardware control remains deterministic and auditable. Figure~\ref{fig:paradigm} summarizes this shift: orchestration moves from fixed human-first workflows to an agent-facing runtime whose hardware interface stays stable. Although this paper focuses on self-driving laboratories, the broader implication is physical AI: PUDA provides an AI-native hardware harness that lets AI systems interact with physical tools through stable commands, live feedback, and auditable records.

This AI-native hardware harness also suggests a longer-term training role. If self-driving laboratories expose command histories, observations, human approvals, rewards, failures, and recoveries as structured records, those records can become interaction data for improving AI models that operate physical laboratories. We envision PUDA as a substrate for a future RL framework in which models learn from physical experimental trajectories while hardware execution remains bounded by deterministic driver methods and human review.

\section{The PUDA Architecture}
\subsection{Design Philosophy}
\paragraph{A command-line interface for physical systems.}
PUDA follows the Unix philosophy of composing small, inspectable commands. A laboratory instrument is not a shell utility, but the interface discipline is similar: expose narrow commands, document their inputs and outputs, return structured results or errors, and let higher-level programs compose those commands into experiments. In practice, a pump, robot arm, potentiostat, camera, or reactor can be represented through the same command-line pattern even though their internal SDKs and vendor APIs are different.

\paragraph{Hardware-agnostic and agent-agnostic by design.}
PUDA is hardware-agnostic because it does not assume one robot brand, instrument type, or laboratory layout. A new device joins PUDA by wrapping its existing API in a reviewed edge driver that exposes atomic commands. Here, atomic means that each command asks the device to do one clearly defined thing, such as moving to a position, dispensing a specified volume, or acquiring a measurement. More complex procedures are built by placing these small commands in sequence within a protocol. PUDA can therefore check and record each step separately, making it clear which action was requested, whether it completed, and where a failure occurred. PUDA is also LLM-agnostic because the runtime does not depend on a specific agent framework or model provider. Any agent, script, scheduler, or human operator that can use a terminal can discover machines, inspect commands, submit protocols, and read results through the same interface \cite{pudadocs}.

\paragraph{A runtime environment, not an orchestrator.}
PUDA does not decide which hypothesis to test or which experiment is scientifically optimal. The agent owns that reasoning. PUDA provides the runtime interface that lets the agent's reasoning touch the physical world: discover available devices, validate the requested protocol, dispatch commands to hardware, observe machine state and outputs, record command responses and data products, and recover provenance after the run. These steps map naturally onto the agent's OODA loop: PUDA helps the agent observe the lab, orient using structured records, decide the next protocol, and act through validated hardware commands while keeping physical execution deterministic and auditable.

\subsection{Layered Overview}
\begin{figure}[t]
\centering
\includegraphics[width=0.92\linewidth]{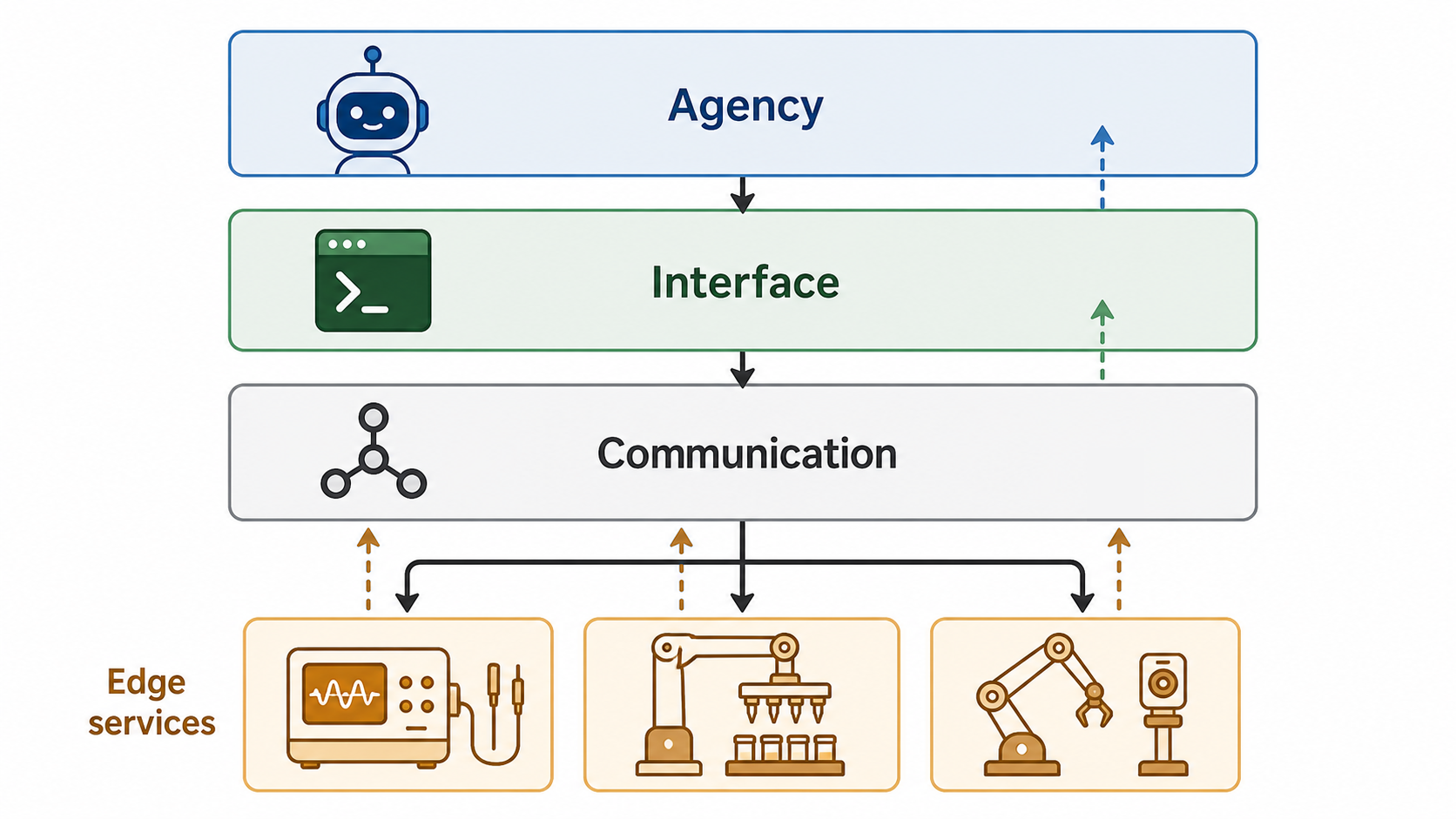}
\caption{PUDA separates the agent, command-line interface, distributed messaging layer, and edge services. Solid arrows show action flow from agent to hardware; dashed arrows show feedback returning as context for the next agent step.}
\label{fig:architecture}
\end{figure}

PUDA separates agentic laboratory operation into four layers, as shown in Figure~\ref{fig:architecture}. The agency layer is the AI agent itself. PUDA does not define the agent's internal planner, memory, or reasoning loop; it gives the agent tools through which it can operate the laboratory. The interface layer is the PUDA command-line interface. This layer provides context to the agent through progressive discovery: the agent can list machines, inspect available commands, validate protocols, read error messages, query records, and navigate toward the next executable action.

The communication layer moves context between the command-line interface and the physical laboratory. PUDA implements this layer with a distributed messaging system based on NATS.io. Commands, telemetry, events, and machine state move through named messaging channels; JetStream adds durable queues and streams for messages that require retention, replay, or coordination across long-running tasks. In this sense, the communication layer is not just transport. It is the path by which live hardware state becomes usable context for the agent.

The edge layer contains the device-local services. Each edge service wraps one instrument's SDK, serial interface, HTTP API, or vendor software behind reviewed driver methods. The edge service receives commands, executes deterministic hardware operations, and publishes live feedback such as heartbeat state, command responses, errors, sensor readings, logs, and results. That feedback returns through the communication and CLI layers, where it becomes the next context the agent uses to observe, orient, decide, and act.

Figure~\ref{fig:architecture} also distinguishes the two directions of control. Solid arrows represent action flow: the agent issues a command or protocol through the PUDA CLI, the interface validates it, the communication layer routes it, and the appropriate edge service dispatches the reviewed hardware command. Dashed arrows represent feedback flow: edge services return telemetry, command responses, errors, logs, and results so the agent can understand what happened and decide the next step. This action-feedback structure is what allows PUDA to support OODA-style agent operation while keeping hardware execution deterministic and auditable.

\subsection{Drivers: Predefined Commands as the Machine Contract}
PUDA makes the driver the boundary between agent reasoning and physical actuation. Each machine is wrapped by a small driver whose public methods define the commands that an agent can call during a run. During an experiment, the agent does not rewrite low-level device-control code. It discovers the available command catalog, selects a reviewed command, supplies structured parameters, and receives a structured response.

The command catalog is generated from the driver itself. Method names, signatures, and docstrings become the information shown by commands such as \texttt{puda machine commands \textless machine\_id\textgreater}. For an agent, this catalog explains what the machine can do and which inputs are required. For an engineer, it is the review surface: public methods are intentional hardware actions, while helper functions, vendor objects, connection details, and unsafe internal operations stay private inside the driver.

PUDA also provides templates for writing new drivers. A driver template gives the developer a standard structure for machine initialization, connection handling, public command methods, state reporting, error handling, and documentation. This makes it easier to wrap a new SDK, serial device, HTTP service, or vendor program without changing the agent-facing runtime. The template does not remove engineering judgment; it makes the machine contract explicit and easier to review.

This boundary does not mean that agents can never help write or improve driver code. A coding agent may update a driver method from human input, debugging feedback, or observations from previous runs. The constraint is temporal and procedural: driver code should not be rewritten while an experiment is actively executing. Once a new or modified driver method has been reviewed, validated, and restarted as part of an edge service, it becomes part of the command catalog that future agents can safely discover and call.

This contract keeps PUDA hardware-agnostic. A liquid handler, robot arm, potentiostat, camera, or reactor may use different SDKs internally, but PUDA only requires a clear set of atomic, documented driver commands. The device-specific implementation stays inside the driver; the agent-facing runtime remains consistent.

\subsection{Edge Services and NATS Transport}
An edge service is the device-local bridge between NATS and a reviewed driver. It runs near the instrument, receives commands addressed to that machine, invokes the corresponding driver method, and returns the result. Hardware-specific SDKs and device-control logic remain at the edge, while the CLI presents agents and operators with a consistent command surface.

Telemetry is the feedback path from the edge service to the CLI and other subscribers. It includes heartbeat and machine state, command status and responses, errors, sensor readings, logs, and data products. Core NATS carries live updates through publish/subscribe subjects, JetStream retains selected events and data for durable consumption or replay, and KV buckets expose the latest machine state and command catalog.

\begin{figure}[t]
\centering
\includegraphics[width=0.88\linewidth]{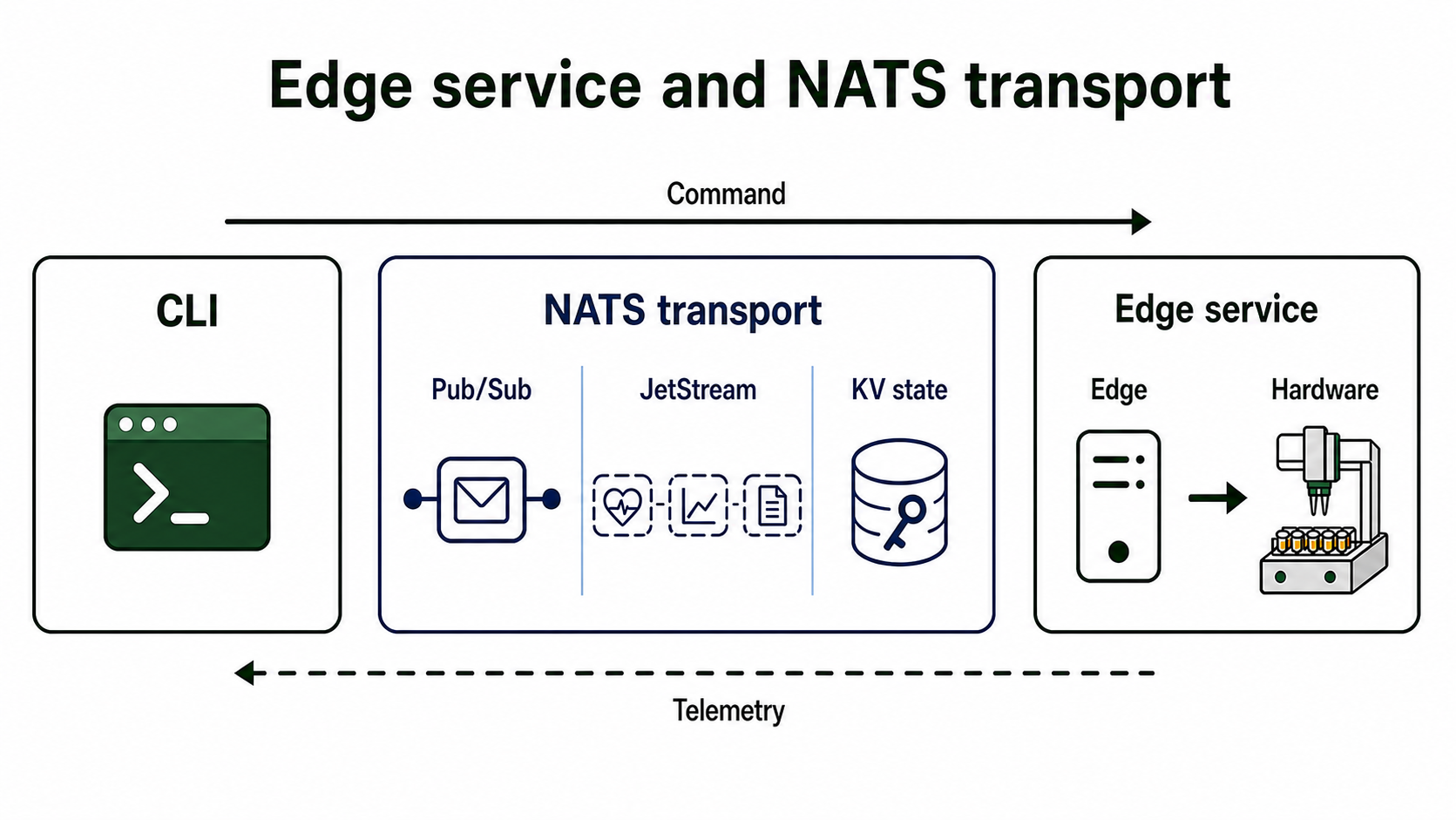}
\caption{Commands move from the CLI through NATS to an edge service, which invokes reviewed hardware operations. Telemetry returns through publish/subscribe for live observation, while JetStream retains selected events and data and KV exposes current machine state.}
\label{fig:transport}
\end{figure}

Figure~\ref{fig:transport} summarizes these two directions: commands move from the CLI through NATS to an edge service and its hardware, while telemetry returns through the same transport as context for the next action. Because publishers and subscribers do not need to run in one process or on one computer, the same pattern supports distributed devices, agents, loggers, and monitors \cite{natsdocs}.

\section{The Command-Line Interface as an Agent Interface}
\begin{figure}[t]
\centering
\includegraphics[width=0.90\linewidth]{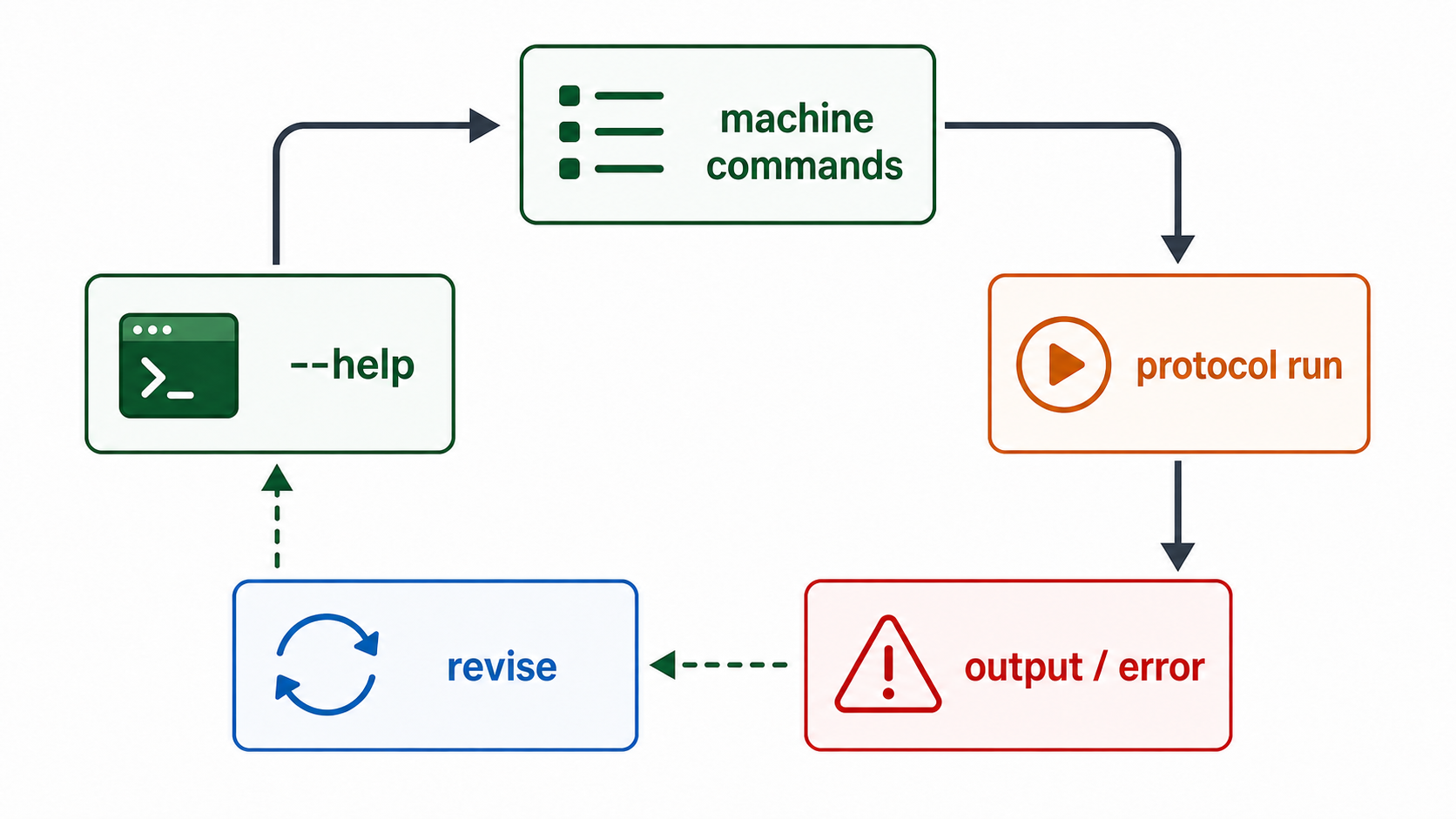}
\caption{CLI interaction as an agent loop: discover commands, inspect contracts, act, and use output or errors to revise the next call.}
\label{fig:cli-loop}
\end{figure}

PUDA exposes the laboratory through a command-line interface because a CLI is compact, explicit, and naturally suited to agentic operations. A GUI is natural for human pointing and clicking, but it often hides state behind screens, widgets, and visual workflows. A CLI turns the same interaction into short commands, structured outputs, error messages, and logs that are token-efficient for language models while still being readable by humans. Figure~\ref{fig:cli-loop} summarizes this loop: the agent discovers the command surface, inspects the contract, acts, and uses output or errors to revise the next call.

The current PUDA CLI is implemented as a Go application and distributed as a binary, giving agents a stable command surface for operating the runtime. Go's cross-compilation support means the same codebase is built and shipped as native binaries for Windows, macOS, and Linux, so agents and human operators get an identical command surface regardless of the host operating system. Through this binary, an agent can discover machines, inspect command catalogs, validate protocols, run experiments, query records, and recover from errors all without loading a graphical interface. This makes the CLI more than a convenience layer: it is the shared operating surface through which agents, scripts, and human operators interact with PUDA.

Because the interface is a standard CLI, PUDA can be integrated with many coding agents, agentic frameworks, and laboratory orchestration systems without custom GUI automation. Coding agents such as Codex, Cursor, and Claude Code can call PUDA commands directly from a terminal or project workspace. Agent frameworks such as OpenClaw or Hermes can wrap the same commands as tools \cite{openclaw,hermesagent}. Scientific co-agent systems such as Google's Co-Scientist and FutureHouse's Robin generate hypotheses, propose experiments, and interpret scientific evidence; PUDA could provide the hardware-facing execution and provenance layer that connects such reasoning systems to physical laboratories \cite{gottweis2025coscientist,ghareeb2025robin}. Laboratory orchestration and agentic-SDL systems such as IvoryOS, HELIOS, and La Agente can use the same CLI boundary as a runtime layer for hardware execution and records \cite{ivoryos2025,helios2026,laagente2026}. In this sense, the CLI makes PUDA composable: it can serve as the hardware-facing command surface beneath different planners, agents, and laboratory software stacks. Listing~\ref{lst:puda-help} shows the compact command surface that an agent can discover before selecting a more specific machine or protocol action.

\begin{lstlisting}[caption={Representative PUDA CLI surface used by agents and humans. Exact subcommands vary by release; each branch exposes further help with \texttt{--help}.},label={lst:puda-help}]
$ puda --help
Available Commands:
  completion  Generate the autocompletion script for the specified shell
  config      Manage PUDA CLI configuration
  db          Database operations
  help        Help about any command
  init        Initialize a new PUDA project
  login       Log in to a PUDA account
  logout      Log out of a PUDA account
  machine     Machine operations
  project     Project operations
  protocol    Protocol operations
  skills      Manage puda agent skills
  update      Update the puda CLI to the latest (or a specific) release
  version     Print the version information

$ puda --version
puda version <installed-release>
\end{lstlisting}

\subsection{Agentic Skills and Progressive Discovery}
PUDA exposes low-level machine commands through the CLI, but most laboratory work happens at a higher level than a single command. For that reason, PUDA is designed to work naturally with agentic skills. A skill can describe how to run a workflow, prepare a protocol, interpret machine state, extract data, or generate a report. In practice, these skills can be Markdown-based workflow templates that tell an agent how to use PUDA commands, what files to inspect, which records to preserve, and how to recover when an experiment fails.

This makes progressive discovery more useful than a static, up-front schema. The agent does not need to load the full laboratory interface at once. It can start from the CLI, discover available machines, inspect command catalogs, generate a protocol, validate a protocol, query the database, and then call a domain-specific skill when the workflow requires more structure. The same pattern can build on agent harnesses such as OpenClaw or Hermes, where memories of past lab workflows, protocol templates, and analysis procedures can guide future orchestration \cite{openclaw,hermesagent}.

PUDA can also coexist with Model Context Protocol (MCP) servers or other schema-first tool interfaces \cite{hou2025mcp}. MCP-style servers are useful when a tool contract should be exposed up front. In practice, PUDA favors the CLI as the primary laboratory interface because it is token-efficient, easy to inspect, and scales through progressive discovery. The agent can ask only for the command branch, machine catalog, state, or report it needs, while prompts, machine operations, state snapshots, measurement results, and reports are stored as records for later agents to reuse.

\clearpage

\subsection{Safety Boundary and Recoverable Operation}
\begin{figure}[t]
\centering
\includegraphics[width=0.90\linewidth]{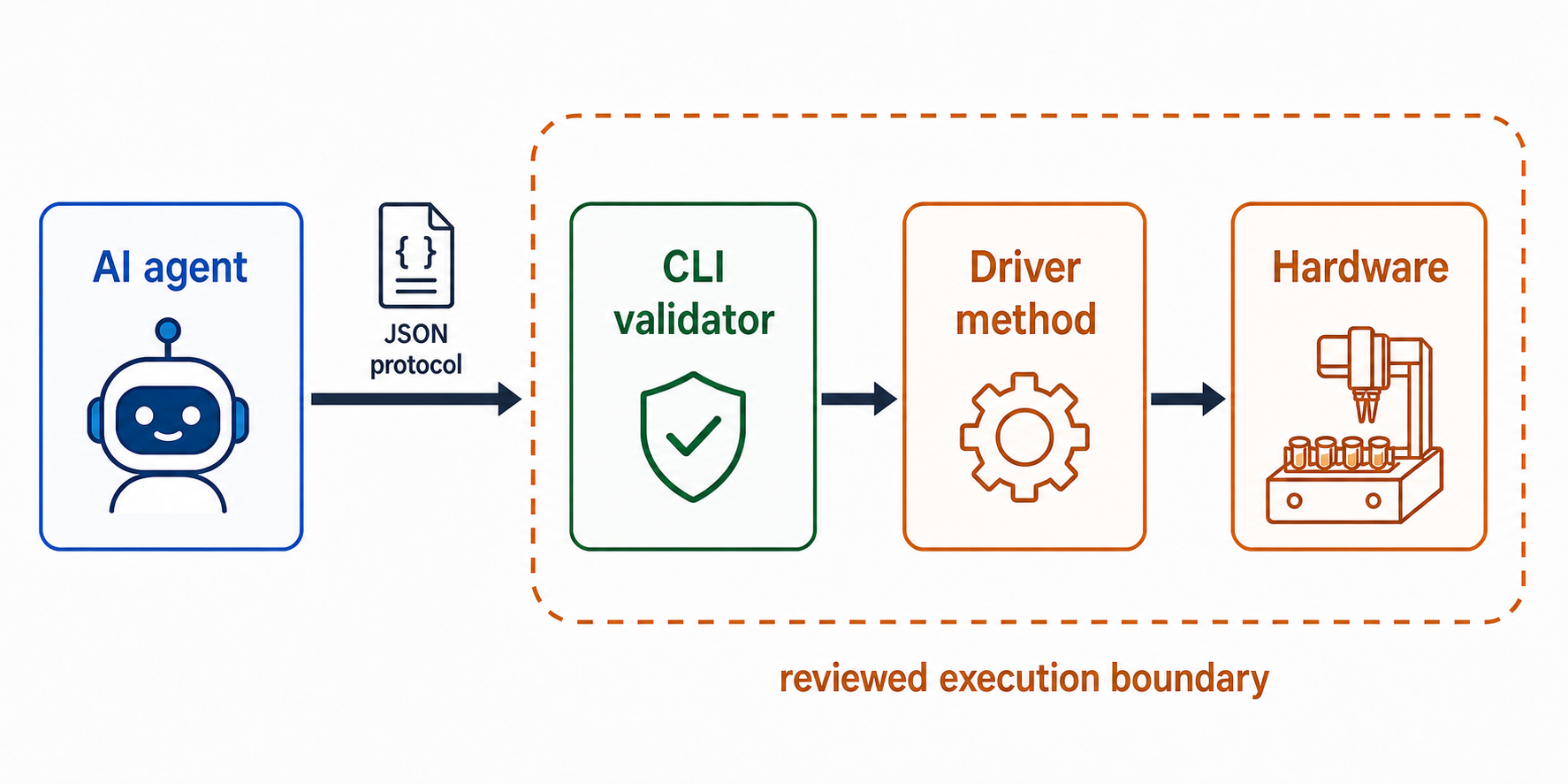}
\caption{PUDA's safety boundary is conventional: agents compose named commands; validation checks protocol structure; predefined driver methods own hardware semantics.}
\label{fig:safety}
\end{figure}

Recent work argues that self-driving laboratories need an autonomy safety harness connecting AI-generated intent to executable experiments, monitored actions, and trustworthy evidence \cite{chen2026safetyharness}. Figure~\ref{fig:cli-loop} also shows why the CLI is useful for safety: action and feedback remain visible at each step. PUDA's safety boundary is simple: agents compose protocols, but reviewed driver methods own hardware actuation. Figure~\ref{fig:safety} makes this driver boundary explicit: the agent prepares a named command or protocol, validation checks its structure, and predefined driver methods hold the hardware semantics. During normal operation, an agent can choose a named command and supply parameters, but it cannot rewrite low-level device-control code during the experiment. Before a command reaches hardware, PUDA can validate the protocol structure, check that the command exists, and route the request only to the intended edge service.

The CLI also makes failure states recoverable. If a device is disconnected, a command is invalid, or an experiment enters an unexpected state, the agent can rediscover available machines, inspect current state, read command errors, reset a device when a reviewed reset command exists, and revise the next protocol. The goal is not to hide errors from the agent, but to return errors as structured context for the next action.

For higher-risk hardware actions, PUDA can use the approval mechanisms already common in coding-agent environments. A coding agent may draft a protocol, explain the intended hardware action, and wait for human approval before dispatching commands that move instruments, apply voltage, heat a reactor, or consume samples. Device-specific safety constraints still belong in the driver, validator, interlocks, lab procedure, or human review.

\subsection{Records, Provenance, and Security}
PUDA records the execution path in local database tables and in structured NATS messages. The CLI initializes a local SQLite schema with \texttt{project}, \texttt{protocol}, \texttt{run}, \texttt{sample}, \texttt{measurement}, and \texttt{command\_log} tables. When \texttt{puda protocol run} is called, PUDA loads the protocol JSON, validates the command structure, stores the protocol, creates a UUID \texttt{run\_id}, sends commands over NATS, and records machine responses in \texttt{command\_log} with \texttt{run\_id}, step number, command name, machine id, command type, JSON payload, and \texttt{created\_at} timestamp \cite{pudadocs}. The NATS message header also carries user id, username, machine id, run id, and UTC timestamp, while each command response carries status, completion time, error code or message, and returned data. Edge services publish live machine state and command catalogs into NATS key-value buckets, so the runtime records both the command history and the current context an agent uses for recovery.

For provenance and security, PUDA keeps experimental records on premises. Each project uses a local SQLite database, and the NATS cluster can run within the laboratory network. An optional logger can copy selected NATS messages to a central PostgreSQL database, creating a shared backup and data pool while retaining the local project record \cite{pudadocs,natsdocs}. The \texttt{puda project hash} command computes a deterministic SHA-256 hash from the recorded project data. If a protocol, run, measurement, or generated data record changes, the hash also changes. UUIDs and timestamps link commands, hardware responses, and resulting data products, making the experimental history traceable.

PUDA data remains on premises regardless of whether the agent uses a cloud-hosted, on-premises, or locally hosted model. Because PUDA is LLM-agnostic, a laboratory can choose any of these model options without moving its databases or runtime services. Only information intentionally sent to a cloud model leaves the laboratory network.

\section{Installation, Operation, and Closed-Loop Use}
\subsection{Deployment Models}
PUDA can run as a single-machine stack or as a distributed runtime across a laboratory network, with the same CLI, NATS transport, edge services, agents, and local records in either case. Release-specific installation steps are maintained at \url{https://docs.puda.co/} \cite{pudadocs}.

A minimal setup can run the AI Agent, PUDA CLI, NATS server, logging service, and edge services all on the same computer. This mode is useful for testing a new machine, reproducing an example workflow, or operating a single instrument during development. For routine laboratory use, PUDA encourages a distributed setup. The NATS server can run on an always-on machine, while edge services run close to their instruments on a Raspberry Pi, other single-board computer, mini PC, Mac mini, or the workstation already connected to the device.

This deployment model keeps the communication layer stable while allowing each instrument to use the operating system, vendor software, and physical connection that fits the hardware. A potentiostat, liquid handler, camera, or robot arm may live on different computers, but all of them appear to the agent through the same PUDA command and record layer.

\subsection{Agent-Assisted Setup}
Setup can also be delegated to a shell-capable coding agent. The operator can point the agent to the PUDA documentation, ask it to install the CLI, configure the NATS address, install machine skills, start an edge service, and verify that machines appear in the command catalog. This is not a separate automation mechanism; it is the same design used during experiments. PUDA exposes files, commands, logs, and errors that an agent can inspect and repair.

Machine skills make this process less fragile. A skill can describe how a device is started, which commands or parameters are safe to call, how to interpret data from responses, where reports are written, and what recovery actions are available. The skill does not replace the driver; it gives the agent operating extra context around the driver.

\subsection{Operating Through PUDA Commands}
After installation, operation starts with discovery. The agent reads the available skills, lists machines, inspects a machine's command catalog, writes a JSON protocol, validates the protocol, and runs it. The result is not just a hardware action; it is a traceable sequence from intent to command response and record. Listing~\ref{lst:protocol-run} gives the shorter operational path, while Table~\ref{tab:trace} summarizes how the same path becomes a record.

\begin{lstlisting}[caption={Representative command path adapted from the PUDA documentation. Exact subcommands and output depend on the installed PUDA release and live machine metadata.},label={lst:protocol-run}]
# Install or update PUDA agent skills
$ puda skills install

# Authenticate the user or agent with the PUDA server
$ puda login --username <username>

# List machines visible to the current PUDA environment
$ puda machine list

# Show callable commands and parameters for a machine
$ puda machine commands <machine_id>

# Check that the protocol structure is valid
$ puda protocol validate -f protocols/example.json

# Dispatch the validated protocol to edge services
$ puda protocol run -f protocols/example.json
\end{lstlisting}

\begin{table}[t]
\centering
\caption{PUDA trace from request to record.}
\label{tab:trace}
\begin{tabularx}{\linewidth}{@{}lXX@{}}
\toprule
Stage & PUDA surface & Artifact \\
\midrule
Intent & User or agent request & Natural-language instruction. \\
Discovery & \texttt{puda machine}; \texttt{puda machine commands \textless id\textgreater} & Live machine IDs and callable methods. \\
Execution & \texttt{puda protocol validate -f}; \texttt{puda protocol run -f} & Validated JSON protocol and run ID. \\
Evidence & Logging service, database, reports, memory & Responses, hashes, report artifacts, and project history. \\
\bottomrule
\end{tabularx}
\end{table}

\subsection{Creating Optimization Loops}
\begin{figure}[t]
\centering
\includegraphics[width=0.88\linewidth]{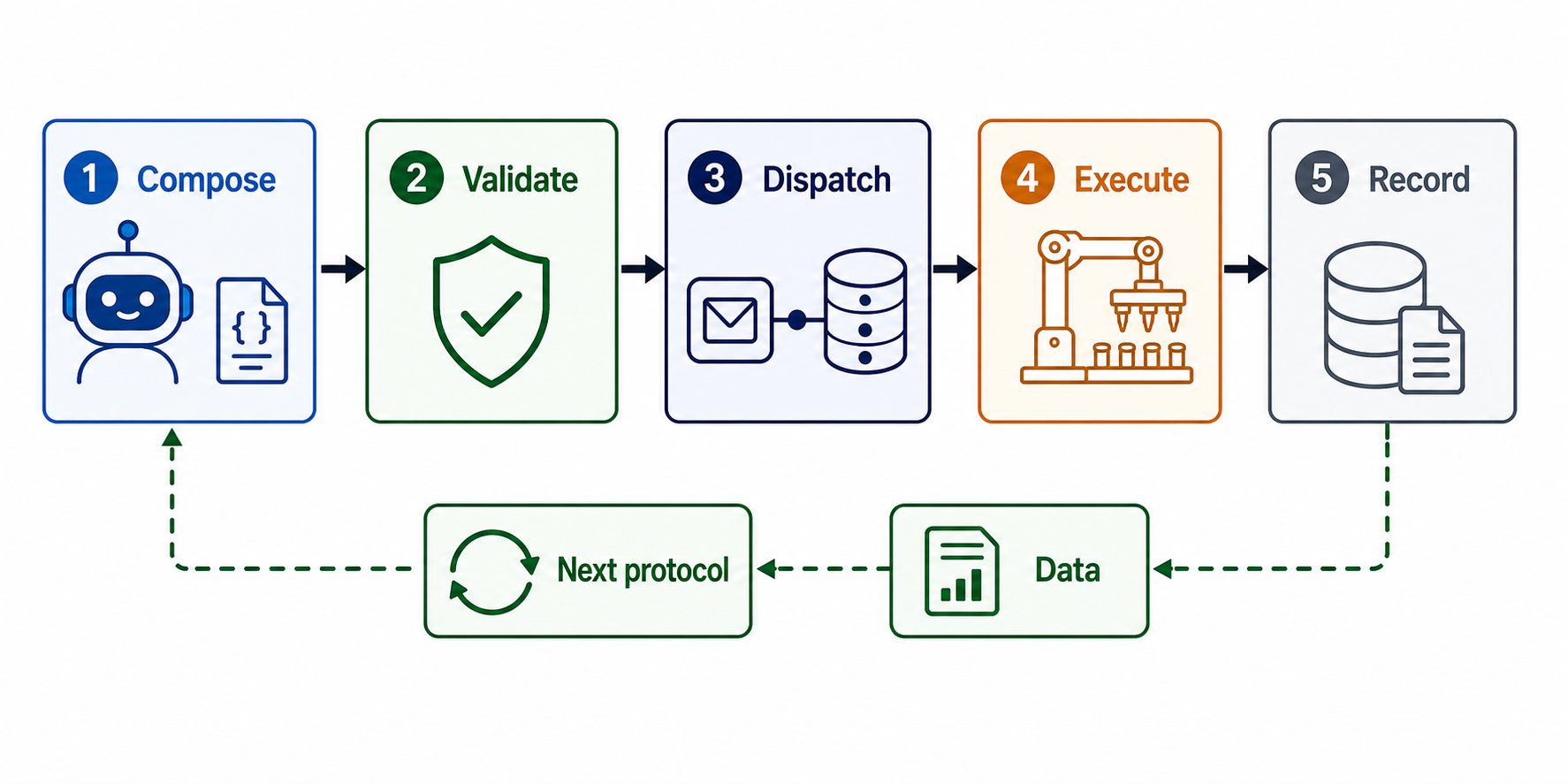}
\caption{PUDA closes the experimental loop without placing workflow logic in the driver: an agent composes the protocol, the CLI validates it, NATS dispatches commands, reviewed drivers execute the physical step, and structured data and provenance inform the next protocol.}
\label{fig:lifecycle}
\end{figure}

PUDA can help close an optimization loop without becoming the optimizer. A policy may be a human-written script, a Bayesian optimizer, a coding agent, or a multi-agent system. PUDA supplies the runtime beneath that policy: discoverable commands, validated dispatch, live feedback, persistent records, and recoverable errors. Figure~\ref{fig:lifecycle} illustrates the generic closed-loop pattern: the policy decides what to try next; PUDA executes the physical step and preserves what happened.

As one example, in a viscosity liquid-handling experiment, the agent can ask PUDA to dispense a volume, read the measured weight, compare it with a target value, and adjust the next dispense volume until the target tolerance is reached. The same logic extends to richer experimental setups: the feedback signal can be a spectrum, electrochemical trace, image-derived color, reactor temperature profile, chromatogram, or multi-instrument report. PUDA keeps the loop executable and auditable while the optimization strategy remains outside the driver.

PUDA, so far, has been deployed across multiple AI4Science laboratory contexts, including inorganic automated sample handling, materials synthesis, and device measurement workflows. These deployments differ in instruments and scientific objectives, but they share the same runtime pattern: an agent or operator discovers the available machine commands, dispatches reviewed hardware actions, observes live feedback, and preserves the resulting records for later analysis.

PUDA is also useful when the loop is not fully autonomous. Because the CLI layer can be coupled to an agentic framework, a researcher can interact with the campaign while it is running: inspect the proposed protocol, review machine state, approve or revise a hardware action, and evaluate the result or reward signal after each run. This human-in-the-loop mode keeps expert judgment inside the experimental loop while PUDA preserves the same command trail, feedback, and records used by fully closed-loop optimization. PUDA is not an agent framework; it is the SDL-facing runtime and data environment that lets one agent or many agents close experimental loops through the same commands, driver methods, streams, and records.

\section{Related Work and Design Position}
PUDA builds on a large body of work in self-driving laboratories, autonomous experimentation, laboratory workflow systems, and model-facing tool interfaces. Reviews and roadmaps have established why closed-loop experimental systems matter for chemistry and materials science \cite{tom2024,abolhasani2023,hase2019}. Demonstrations such as the mobile robotic chemist, thin-film self-driving laboratories, A-Lab, autonomous data-driven synthesis, and LLM-driven autonomous chemistry show that physical experimentation can be coupled to optimization, machine-learning, and agentic planning loops \cite{burger2020robotic,macleod2020thinfilm,szymanski2023alab,dai2024organic,boiko2023coscientist}. Metrics work further clarifies how such systems should be evaluated \cite{volk2024metrics}. PUDA does not claim a new autonomous-discovery result. Its contribution is the runtime environment that lets agents operate such loops through a stable, inspectable hardware interface.

Workflow-centered SDL and orchestration platforms solve a complementary problem. ChemOS, HELAO, AlabOS, IvoryOS, HELIOS, La Agente, ARES OS, Bluesky, NIMS-OS, Chemputer, and ESCALATE organize devices, recipes, schedulers, data acquisition, optimization loops, or laboratory automation interfaces \cite{chemos2024,helao2022,alabos2024,ivoryos2025,helios2026,laagente2026,sloan2026ares,bluesky2019,mehr2020chemputer,nimsos2023,pendleton2019escalate}. Recent systems such as HELIOS and La Agente make the agent-facing direction more explicit by combining optimization with autonomous orchestration. PUDA differs in design emphasis. Rather than making a GUI, drag-and-drop graph, or fixed recipe editor the center of orchestration, PUDA is headless and AI-first: the agent is expected to plan, revise, and recover while PUDA keeps the executable laboratory surface stable.

SiLA 2 is best viewed separately as a laboratory communication and automation standard \cite{sila}. PUDA uses a different runtime communication protocol internally: NATS/JetStream provides publish/subscribe messaging, durable streams, and multi-agent coordination for live SDL operation. This choice does not exclude SiLA 2. PUDA supports SiLA 2-compatible instruments at the edge layer by writing PUDA drivers that wrap around the gRPC API layer, so an agent can still discover and call them through the same CLI, protocol, and record surface.

Model-facing tool interfaces, including MCP-style systems, make tool contracts visible to AI agents \cite{hou2025mcp}. PUDA shares the goal of explicit, inspectable actions, but physical laboratories add constraints that pure software tools often do not face: hardware state changes over time, device errors can be physical, actions may consume samples, and unsafe commands must be blocked before they reach an instrument. PUDA therefore favors progressive discovery through a CLI, live machine catalogs, NATS/JetStream transport, and persistent command records rather than relying only on a static tool schema.

The central design boundary is between orchestration and actuation. An orchestrator decides which tasks should run, in what order, with which dependencies, branches, schedules, and recovery policies; an AI agent can make the same campaign-level decisions dynamically. PUDA does not define this control flow. It exposes reviewed device commands, validates and dispatches them through edge drivers, and records command responses, telemetry, logs, data products, and reports. An optimizer selects what to try next, an orchestrator coordinates how the experiment proceeds, and PUDA provides the hardware-facing runtime beneath both.

PUDA also inherits the ordinary responsibilities of physical automation. It does not by itself choose the scientific objective, calibrate an instrument, certify chemistry safety, or guarantee that every possible hardware command is safe. Those responsibilities remain in driver range checks, validators, interlocks, laboratory procedures, human approval, and domain-specific skills. The claim is narrower and practical: PUDA makes agent-operated laboratory work callable, recoverable, deterministic at the hardware boundary, and auditable after the run.

\section{Conclusions and Future Work}
Self-driving laboratories need more than automated instruments and optimization algorithms. As AI agents become better at planning, tool use, and recovery, laboratories need a runtime environment that those agents can operate directly without turning every experiment into a fixed GUI workflow or bespoke script. PUDA addresses this interface problem by making the self-driving laboratory headless, discoverable, and auditable.

The contribution is the hardware harness itself: a shared CLI, progressive command discovery, NATS/JetStream-backed edge services, reviewed driver methods, deterministic and atomic hardware actions, and persistent records of commands, state, data products, and reports. Agents decide what experiment to run next; PUDA controls how validated commands reach devices and how evidence is recovered after the run.

Future work will include adding a PUDA simulator that lets agents validate and debug physical protocols in a digital-twin setting, following MatteriX's use of NVIDIA Isaac Sim and Isaac Lab for robotics-assisted chemistry laboratory simulation~\cite{matterix2026}. Longer term, PUDA could provide structured physical interaction trajectories for an RL framework, where AI models learn from experimental actions, observations, rewards, failures, and human review while hardware execution remains bounded by reviewed drivers.

\section*{Acknowledgements}
This research is supported by the National Research Foundation, Prime Minister's Office, Singapore under its Campus for Research Excellence and Technological Enterprise (CREATE) programme.

\end{document}